\documentclass[twocolumn,floatfix,showpacs,eqsecnum,aps,prb,10pt]{revtex4-1}
\pdfoutput=1
\usepackage[colorlinks=true, citecolor=blue]{hyperref}
\usepackage{graphicx}
\usepackage{amsmath}
\usepackage{amssymb}
\usepackage{bm}
\usepackage{color}
\usepackage{multirow}
\DeclareMathOperator{\Pf}{Pf}
\DeclareMathOperator{\tr}{tr}
\DeclareMathOperator{\imag}{Im}
\DeclareMathOperator{\sign}{sign}
\newcommand\pdag{{\hphantom\dag}}

\begin{document}

\title{Scattering theory of topological insulators and superconductors}
\author{I. C. Fulga}
\affiliation{Instituut-Lorentz, Universiteit Leiden, P.O. Box 9506, 2300 RA Leiden, The Netherlands}
\author{F. Hassler}
\affiliation{Instituut-Lorentz, Universiteit Leiden, P.O. Box 9506, 2300 RA Leiden, The Netherlands}
\author{A. R. Akhmerov}
\affiliation{Instituut-Lorentz, Universiteit Leiden, P.O. Box 9506, 2300 RA Leiden, The Netherlands}
\date{July 2011}
\begin{abstract}
The topological invariant of a topological insulator (or superconductor)
is given by the number of symmetry-protected edge states present at the
Fermi level.  Despite this fact, established expressions for the topological
invariant require knowledge of all states below the Fermi energy. Here,
we propose a way to calculate the topological invariant employing solely
its scattering matrix at the Fermi level without knowledge of the full
spectrum. Since the approach based on scattering matrices requires much less
information than the Hamiltonian-based approaches (surface versus bulk),
it is numerically more efficient. In particular, is better-suited for
studying disordered systems. Moreover, it directly connects the topological
invariant to transport properties potentially providing a new way to probe
topological phases. 
\end{abstract}

\maketitle

\section{Introduction}
\label{intro}

Given a Hamiltonian $H(\bm{k})$ of a band insulator or a superconductor
and its symmetries as a function of the momentum $\bm{k}$ in $d$-spatial
dimensions, a topological invariant $\mathcal{Q}(H)$ can be defined. It
counts the number of surface states insensitive to disorder which are present
at an interface between the system and the vacuum. In each spatial dimension
exactly 5 out of 10 Altland-Zirnbauer symmetry classes (distinguished by
time-reversal symmetry $\mathcal{T}$, particle-hole symmetry $\mathcal{P}$,
and chiral/sub-lattice symmetry $\mathcal{C}$)\cite{Alt97} allow for a
nontrivial topological invariant.\cite{Kit09, Sch09}

The evaluation of the topological invariant conventionally involves an
integral over a $d$-dimensional Brillouin zone of some function of the
Hamiltonian. Recently, various approximations to the topological invariant
have been developed which require only the knowledge of eigenvalues and
eigenvectors of the Hamiltonian at one point in momentum space (rather
than in the entire Brillouin zone).\cite{Lor10,Rin11,Has11}

Despite the fact that these approximations are more efficient, we argue that
they do not use one important property of a topological invariant.  By
definition, the topological invariant describes the properties of the system
at the Fermi level, namely the number of edge states.  This observation
suggests that it should be possible to calculate the topological invariant
without knowing the full spectrum of the Hamiltonian, but rather calculating
only properties of the system at its Fermi energy.  For one-dimensional (1D)
systems, this was demonstrated in Ref.~\onlinecite{Ful11}.  Here, we show that
\textit{for any dimensionality} the topological invariant can be obtained from
the scattering matrix of the system at the Fermi level.

Our results offer two benefits.  Firstly, since the scattering matrix contains
less degrees of freedom than the Hamiltonian, the computation of the
topological invariant is much more efficient.  Secondly, the scattering matrix
relates the topological invariant to transport properties, suggesting ways to
probe the topological phase by electrical or thermal conduction
measurements.\cite{Akh11,Wim11-1}

The approach is based on dimensional reduction: We relate the scattering
matrix in $d$ dimensions to a Hamiltonian in $d-1$ dimensions.  Our scheme
of dimensional reduction does not preserve the symmetry, unlike the field
theory based scheme of Ref.~\onlinecite{Qi08}. Instead our dimensional
reduction preserves the topological invariant, similarly to the dimensional
reduction of clean Dirac-like Hamiltonians of Ref.~\onlinecite{Ryu10}.

In the remainder of the introduction we first illustrate our approach
by revisiting the familiar example of the integer quantum Hall effect.
Subsequently, we present a brief outline of the paper.

\subsection{Dimensional reduction in the quantum Hall effect}

A 2D system exhibiting the integer quantum Hall effect is a topological
insulator in the symmetry class A (all symmetries broken). It is
characterized by a quantized transverse conductance $\sigma_{xy} =
n g_0$ with $n\in\mathbb{Z}$ and $g_0 = e^2/h$.  The quantum number
$n$ is a topological invariant (the so-called Chern number) of the
Hamiltonian.\cite{Tho82} It equals the number of protected chiral edge
states at the Fermi level, each of which contributes $e^2/h$ to the
transverse conductance.\cite{Hal82,But88}

Charge pumping provides an alternative way to relate the topological
invariant to a quantized transport property: inserting a flux
quantum inside a quantum Hall sample rolled-up to a cylinder adiabatically
pumps $n$ electrons across the sample.\cite{Lau81} There exists a scattering
matrix formulation of charge pumping,\cite{Btt94, Bro98} which allows to
express pumped charge per cycle (in units of $e$),
\begin{equation}
\label{eq:pumping}
\mathcal{Q} = \frac{1}{2\pi i} \int_0^{2\pi} \!\! 
d\varphi \,\frac{d}{d\varphi} \log \det r(\varphi),
\end{equation}
through the flux dependence of the reflection block $r(\varphi)$ of the
scattering matrix of one lead.\cite{Bru09} Here $\varphi$ denotes the
dimensionless flux $\Phi= \hbar \varphi/e$ and the system is assumed to be
insulating such that the reflection matrix $r(\varphi)$ is unitary.
Equation~\eqref{eq:pumping} is nothing but the winding number of $\det
r(\varphi)$ when $\varphi$ is varied from $0$ to $2\pi$, which is a
topological invariant.

The winding number occurs in a different context in the theory of topological
insulators. The topological invariant of a one-dimensional Hamiltonian
$H(k)$ with chiral/sub-lattice symmetry
\begin{equation}
 H(k) = 
\begin{pmatrix}
 0 & h(k) \\
h^\dagger(k) &0
\end{pmatrix},
\end{equation}
is expressed via the winding number given by\cite{Zak89, Ryu02}
\begin{equation}
\mathcal{Q}(H) = \frac{1}{2\pi i} \int_0^{2\pi} \!\! 
dk \,\frac{d}{dk} \log \det h(k).
\end{equation}
Here momentum $k$ is measured in units of $\hbar/a$, with $a$ the
lattice constant.  We see that upon the identification $h \equiv r$ and $k
\equiv \varphi$ we are able to express the topological invariant in a 2D
system without any symmetries as the topological invariant of an effective
Hamiltonian in 1D with chiral symmetry. We will show that a similar dimensional
reduction applies to all topological invariants in all dimensions.

\subsection{Outline of the paper}

As a prerequisite for the dimensional reduction, we have to open up
the system to obtain a scattering matrix from a given Hamiltonian.
Section~\ref{smatrix} explains how this can be done.  This section may
be skipped on first reading. The dimensional reduction proceeds along the
following lines: First we form out of a scattering matrix $S$ a reflection
block $r(\bm{k})$ from one surface of the system, when all the dimensions
except one are closed by twisted periodic boundary conditions. Then, the
effective Hamiltonian $H_{d-1}(\bm{k})$ in one dimension lower is defined
according to the simple rule
\begin{subequations}
\begin{align}
 &H_{d-1}(\bm{k}) \equiv r(\bm{k}), \textrm{\quad with chiral symmetry,}\\
&H_{d-1}(\bm{k}) \equiv
\begin{pmatrix}
 0 & r(\bm{k})\\
r^\dagger(\bm{k}) & 0
\end{pmatrix}, \textrm{\quad without chiral symmetry.}
\end{align}
\end{subequations}
In Sec.~\ref{roll} we show how to evaluate $r(\bm{k})$ given the scattering
matrix $S$ of the initial system and prove that the reduced Hamiltonian
$H_{d-1}$ has the same topological invariant as the original $H$, i.e.
$\mathcal{Q}(H_{d-1}) = \mathcal{Q}(H)$.

After the general proof we turn to the particular ways to evaluate
the topological invariant in 1--3 dimensions in Sec.~\ref{charge}. In
1D we show that our expressions coincide with the ones derived in
Ref.~\onlinecite{Ful11} in a different way, without using dimensional
reduction. For 2D we formulate the evaluation of the topological invariant
as a generalized eigenvalue problem. For 3D topological insulators in
class AII the topological invariant reduces to a product of 2D invariants,
while the other symmetry classes require usage of a Bott index.\cite{Has11}
We also mention how weak topological invariants fit into our approach.

We consider the numerical efficiency of our method and show examples of its
application in Sec.~\ref{numerics}. We also compare the finite size effects
of different approximations to the topological invariant, and introduce the
`fingerprint' of phase transitions between different topological phases
in 2D.  Finally, we conclude in Sec.~\ref{conclusion}. 

\section{Scattering matrix from a Hamiltonian}
\label{smatrix}

This section contains the necessary preliminaries: the definition of
scattering matrix and a proof that the shape of the Fermi surface can be
calculated from the scattering matrix.

While the formulas in this section are needed for the actual implementation of our method of dimensional reduction, the method itself can be understood without them. This section can thus be skipped at first reading.

\begin{figure}[tb]
\includegraphics[width=.7\linewidth]{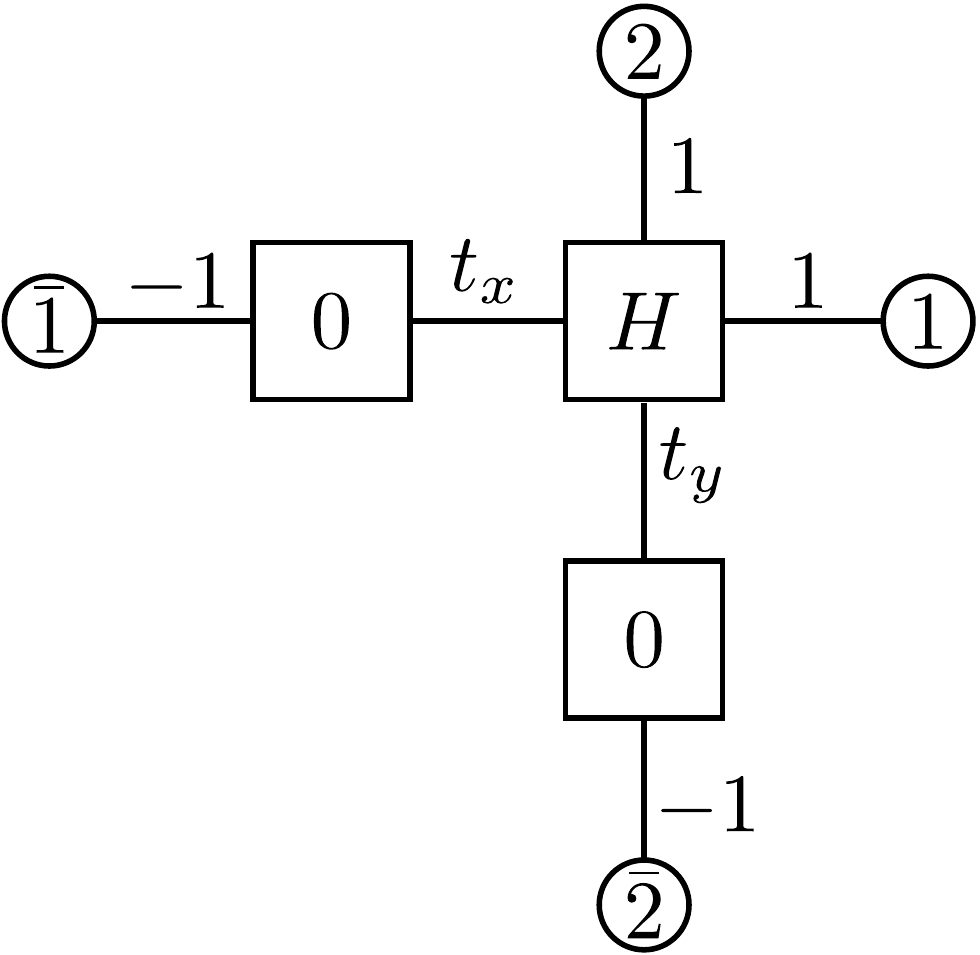}
\caption{%
Sketch of the tight binding model used to attach leads in order to open-up
the Hamiltonian $H(\bm{k})$ of \eqref{eq:ham}. In 2D we introduce four
leads shown as circles labeled by $1$, $\bar1$, $2$, and $\bar2$. The
on-site terms (boxes) are connected by hoppings (lines). The additional
trivial hoppings $1$ and $-1$ are introduced such that the lead properties
drop out when twisted periodic boundary conditions are applied. For the
Mahaux-Weidenm\"uller formula \eqref{eq:weidenmuller}, the three nodes
form the on-site Hamiltonian $\tilde{H}$ which is then connected via the
trivial hoppings to ideal leads.
}\label{fig:scattering}
\end{figure}

Any Hamiltonian $H(\bm{k})$ of a translationally invariant system with a
finite range hopping can be brought to the tight-binding form by choosing
a sufficiently large unit cell
\begin{equation}\label{eq:ham}
 H(\bm{k}) = H + \sum_{i=0}^{d} t_i e^{i k_i} + \sum_{i=0}^{d} t_i^\dagger e^{-i k_i}.
\end{equation}
Here $\bm{k}$ is a $d$-dimensional vector of Bloch momenta, $H$ is the
on-site Hamiltonian, and $t_i$ are the hoppings in positive $i$-direction. We
start our consideration from opening the system and attaching $2d$
fictitious leads to it. First we attach $d$ sites to the original
system without on-site Hamiltonian, and connect them with hoppings $t_i$
to the system. The Hamiltonian of this `unfolded' system becomes
\begin{gather} \label{eq:onsite_ham}
 \tilde{H}=
\begin{pmatrix}
 H & t^\dagger\\
 t & 0
\end{pmatrix},\\
t = (t_1, t_2, \ldots t_d)^T.
\end{gather}
In the next step we attach the fictitious leads to the unfolded system, as
illustrated in Fig.~\ref{fig:scattering} for the case of two dimensions. The
hopping to the leads in positive $i$-direction is chosen to be equal to
$+1$, and in the negative $i$-direction to be equal to $-1$.

We are now ready to construct the scattering matrix of the open system by
using the Mahaux-Weidenm\"uller formula\cite{Mah} (see also Appendix \ref{app:symmetries})
\begin{equation}\label{eq:weidenmuller}
  S= 1 + 2\pi i W^\dag (\tilde{H} - i \pi W W^\dag)^{-1} W.
\end{equation}
The coupling $W$ between the lead and the system is equal to $w\sqrt{\rho}$,
with $w$ the hopping from the lead to the system, and $\rho$ the density
of states in the lead. We choose $\rho = 1/w \pi$, such that
\begin{equation}\label{eq:weiden}
 W = \frac{1}{\sqrt{\pi}}\begin{pmatrix}
  1 &  0 & 1 & 0 & \cdots & 1 & 0 \\
  0 & -1 & 0 & 0 &        & 0  & 0  \\
  0 & 0 &  0 & -1 &       & 0  & 0  \\
  \vdots &   & &  & \ddots&    & \vdots \\
  0 & 0& 0 & 0& \cdots & 0 & -1
\end{pmatrix};
\end{equation}
here, we have set $w=1$ for convenience.
The values of hopping and the lead density of states are chosen such that
in the process of rolling-up, the fictitious leads drop out.

The scattering matrix \eqref{eq:weidenmuller} relates the incoming states
in the leads to the outgoing ones:
\begin{equation}\label{eq:scattering_problem}
 \begin{pmatrix}
  \psi_{1}\\
  \psi_{\bar1}\\
  \psi_{2}\\
  \psi_{\bar2}\\
  \vdots\\
  \psi_{d}\\
  \psi_{\bar{d}}  
 \end{pmatrix}_{\text{out}} = S
 \begin{pmatrix}
  \psi_{1}\\
  \psi_{\bar1}\\
  \psi_{2}\\
  \psi_{\bar2}\\
  \vdots\\
  \psi_{d}\\
  \psi_{\bar{d}}  
 \end{pmatrix}_{\text{in}}.
\end{equation}
To prove that the scattering matrix contains all of the information about the
Fermi level at energy $E_F=0$, we impose twisted periodic boundary conditions
on the scattering states:
\begin{equation}\label{eq:twist}
 \begin{pmatrix}
  \psi_{1}\\
  \psi_{\bar1}\\
  \psi_{2}\\
  \psi_{\bar2}\\
  \vdots\\
  \psi_{d}\\
  \psi_{\bar{d}}  
 \end{pmatrix}_{\text{in}} = Z(\bm{k})
 \begin{pmatrix}
  \psi_{1}\\
  \psi_{\bar1}\\
  \psi_{2}\\
  \psi_{\bar2}\\
  \vdots\\
  \psi_{d}\\
  \psi_{\bar{d}}  
 \end{pmatrix}_{\text{out}},
\end{equation}
with the twist matrix $Z(\bm{k})$ given by
\begin{equation}\label{eq:zmatrix}
Z(\bm{k}) \equiv 
\begin{pmatrix}
  0    & e^{ik_1} &   0     &     \cdots     &  0    \\
  e^{-ik_1} & 0   &        &          &    \vdots  \\
   0      &     & \ddots &          &     0 \\
   \vdots      &     &        & 0        &  e^{ik_d} \\
  0      &   \cdots  &  0      &  e^{-ik_d} & 0
\end{pmatrix}.
\end{equation}
We show that Eqs.~\eqref{eq:scattering_problem} and \eqref{eq:twist} have a
solution for a given $\bm{k}$ if and only if the equation $H(\bm{k})\psi =
0$ has a nontrivial solution. The condition for the nontrivial solution
of Eqs.~\eqref{eq:scattering_problem} and \eqref{eq:twist} to exist is
\begin{equation}
 \det[S-Z(\bm{k})] = 0.
\end{equation}
Performing block-wise inversion of $\tilde{H} - i \pi W W^\dag$ yields
\begin{gather}
 S = 1+2iW^\dagger\begin{pmatrix}
           J & -iJt^\dagger \\
           -itJ & i-tJt^\dagger
          \end{pmatrix} W,\\
J=(H_0-id -it^\dagger t)^{-1}.
\end{gather}
We simplify this expression further by noting that
\begin{align}
S&= \gamma_z + 2i \gamma_z U^\dagger J U,&
U &= \begin{pmatrix}1& -it^\dagger\end{pmatrix} W,
\end{align}
with $\gamma_z$ the third Pauli matrix in the direction space. We now write
\begin{align}
 \det[S-Z(\bm{k})]&=\det\left[1 + \gamma_z Z(\bm{k})+2i U^\dagger J U\right] 
 \\&= 
\det J \det[1+\gamma_z Z(\bm{k})] \nonumber\\
&\quad\times\det\left(J^{-1} + 2i U[1+\gamma_z
Z(\bm{k})] U^\dagger \right) \nonumber\\&=
\det J \det[1+\gamma_z Z(\bm{k})] \det H(\bm{k}). \nonumber
\end{align}
Since both $J$ and $1+\gamma_z Z(\bm{k})$ are nonsingular, the last
identity means that $\det[S - Z(\bm{k})]$ and $\det H(\bm{k})$ can only
be zero simultaneously, which is what we set out to prove.

This proof shows that the Fermi surfaces as defined by the
original Hamiltonian and the scattering matrix are identical.  This is the
reason why it is at all possible to determine the topological invariant using
solely the scattering matrix $S$.  Even though the scattering matrix only
describes scattering at the Fermi level, it contains information about the
complete Brillouin zone, and thus cannot be obtained from a long wavelength or
low energy expansion of the Hamiltonian, but requires the complete
Hamiltonian.  Note however that the scattering matrix at a single energy
contains less information about the system than the Hamiltonian: in order to
determine the Hamiltonian from the scattering matrix, the inverse scattering
problem has to be solved which requires knowledge of the scattering matrix at
all the energies.

The size of the scattering matrix \eqref{eq:weidenmuller} is $2d$-times
larger than the size of Hamiltonian. However, if the Hamiltonian is local
on a large $d$-dimensional lattice with size $L^d$, the hoppings $t_i$
are very sparse. This allows to efficiently eliminate all of the modes
except the ones that are coupled to the hoppings. The resulting scattering
matrix is of size $2dL^{d-1}$, and accordingly for large systems it is a
dense matrix of much smaller dimensions than the Hamiltonian.

\section{Dimensional reduction}
\label{roll}

The aim of this section is to provide a route to the topological
classification of scattering matrices by elimination of one spatial
dimensions.  This approach of dimensional reduction is inspired by the
transport properties of topological systems.  When applied to 1D systems it
reproduces the results of Ref.~\onlinecite{Ful11}, and in quantum Hall systems
it reproduces the relation between adiabatic pumping and the Chern number of
Refs.~\onlinecite{Lau81, Bru09}.

We begin from substituting the first $2(d-1)$ equations from \eqref{eq:twist}
into \eqref{eq:scattering_problem}. This is equivalent to applying twisted
periodic boundary conditions to all of the dimensions except the last one,
which is left open. Then we study the reflection from the $d$-direction
back onto itself. The reflection is given by
\begin{gather}
\psi_{d\textrm{,out}}
= r(\bm{k})
\psi_{d \textrm{,in}},\\
r(\bm{k}) = D - C [A - Z_{d-1}(\bm{k})]^{-1} B, \label{eq:rdef}
\end{gather}
with $Z_{d-1}$ given by Eq.~\eqref{eq:zmatrix} in $d-1$ dimensions.  
The matrices $A$, $B$, $C$, and $D$ are sub-blocks of $S$ given by
\begin{align}
  A&= \begin{pmatrix} S_{1,1} & \cdots & S_{1,\overline{d-1}} \\
    \vdots & \ddots & \vdots \\
    S_{\overline{d-1},1} & \cdots & S_{\overline{d-1},\overline{d-1}}
  \end{pmatrix}, & B=&\begin{pmatrix} S_{1,d} \\ \vdots \\ S_{\overline{d-1},d}
  \end{pmatrix}, \nonumber\\
  C&= \begin{pmatrix} S_{d,1} &\cdots & S_{d,\overline{d-1}}
  \end{pmatrix}, & D&=\begin{pmatrix}S_{d,d}\end{pmatrix}.
\end{align}

To study topological properties of $r(\bm{k})$ we construct an effective
Hamiltonian $H_{d-1}(\bm{k})$ which has band gap closings whenever
$r(\bm{k})$ has zero eigenvalues. In classes possessing chiral symmetry
one may choose a basis such that $r(\bm{k}) = r^\dagger(\bm{k})$. If
chiral symmetry is absent, there is no Hermiticity condition on $r$, so
we double the degrees of freedom to construct a single Hermitian matrix
out of a complex one. The effective Hamiltonian is then given by
\begin{subequations}
\label{eq:reducedham}
\begin{align}
 &H_{d-1}(\bm{k}) \equiv r(\bm{k}), \textrm{\quad with chiral symmetry,}\\
&H_{d-1}(\bm{k}) \equiv
\begin{pmatrix}
 0 & r(\bm{k})\\
r^\dagger(\bm{k}) & 0
\end{pmatrix}, \textrm{\quad without chiral symmetry.}\label{eq:reducedham_chiral}
\end{align}
\end{subequations}
It is straightforward to verify that in both cases the Hamiltonian
$H_{d-1}(\bm{k})$ has band gap closings simultaneously with the appearance
of vanishing eigenvalues of $r(\bm{k})$.

If $r(\bm{k})$ has chiral symmetry,
$H_{d-1}(\bm{k})$ does not have it. On the other hand, if $r(\bm{k})$
has no chiral symmetry, then
\begin{equation}
H_{d-1}(\bm{k}) = - \tau_z H_{d-1}(\bm{k}) \tau_z,
\end{equation}
with $\tau_z$ the third Pauli matrix in the space of the doubled degrees
of freedom. This means that in that case $H_{d-1}(\bm{k})$ acquires chiral
symmetry.

\begin{figure*}[tb]
\centerline{\includegraphics[width=0.9\linewidth]{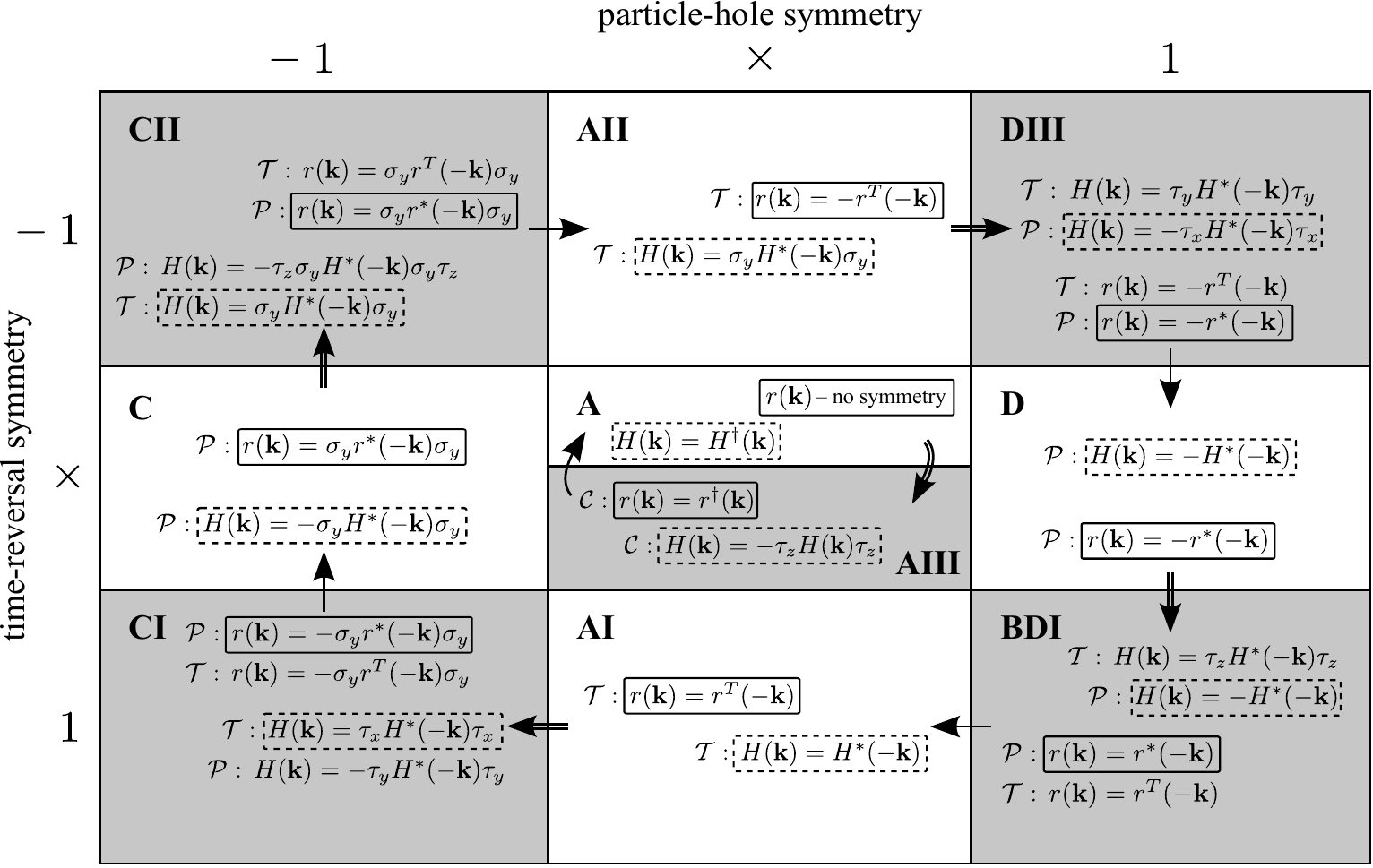}}
\caption{%
Symmetry properties of $r(\bm{k})$ and $H(\bm{k})$ in the ten symmetry
classes.  Time-reversal symmetry is denoted by $\mathcal{T}$, particle-hole
symmetry by $\mathcal{P}$.  The signs at the top and left of the table denote
either the absence ($\times$) of a corresponding symmetry, or the value of the
squared symmetry operator.  The entries of the table with a gray background
have an additional chiral symmetry $\mathcal{C}$, which always has the form
shown in the AIII entry of the table.  In particular, we always chose a basis
such that $r(\bm{k}) = r^\dag (\bm{k})$ in the chiral symmetry classes.  The
way symmetry classes transform under our definition of $H_{d-1}$,
cf.~\eqref{eq:reducedham}, is denoted by the arrows; the double arrow implies
a doubling of degrees of freedom as in Eq.~\eqref{eq:reducedham_chiral}.
Going along an arrow, the symmetry of the reflection block $r(\bf{k})$ (marked
by a solid box) transforms into the symmetry of the reduced Hamiltonian
(marked by a dashed box).  In the chiral classes, there is an additional
symmetry (not marked by a box) which can be obtained from the other by
combining it with the chiral symmetry, $H(\bm{k}) = -\tau_z H(\bm{k}) \tau_z$
and $r(\bm{k}) = r^\dag (\bm{k})$, respectively.
}\label{fig:symmetries}
\end{figure*}

The way in which the dimensional reduction changes the symmetry class 
is summarized in Fig.~\ref{fig:symmetries}. The
transformation of symmetries of $r(\bm{k})$ into symmetries of
$H_{d-1}(\bm{k})$ is straightforward in all of the cases, except the
time-reversal symmetry in symmetry classes AII and AI. There we have
$r(\bm{k}) = \pm r^T(-\bm{k})$, and hence
\begin{multline}
 H_{d-1}(\bm{k}) \equiv 
\begin{pmatrix}
0 & r(\bm{k})\\
r^\dagger(\bm{k}) & 0                    
\end{pmatrix} = \\
\begin{pmatrix}
0 & \pm r(-\bm{k})^T\\
\pm [r^T(-\bm{k})]^\dagger & 0                    
\end{pmatrix} = \pm \tau_x H_{d-1}^{*}(-\bm{k})\tau_x. \label{eq:hfromr}
\end{multline}
The details of the symmetry properties of $r$ and $H$, as well as the relations
between these symmetries are given in App.~\ref{app:symmetries}.

The way the symmetry class of the $d$-dimensional Hamiltonian transforms into
the symmetry class of $H_{d-1}(\bm{k})$ expresses the Bott periodicity of
the topological classification of symmetry classes.\cite{Kit09} Namely,
symmetry classes A and AIII transform into each other, and the other
8 classes with anti-unitary symmetries are shifted by one, as shown in
Table~\ref{tab:az}. This reproduces the natural succession of symmetry classes that appears in the context of symmetry breaking\cite{Sto10} (see also Appendix \ref{app:symmetries}). The combined effects of the change in dimensionality and
in symmetry class is that the Hamiltonians $H(\bm{k})$ and $H_{d-1}(\bm{k})$
have the same topological classification.

\begin{table}[tb]
\includegraphics{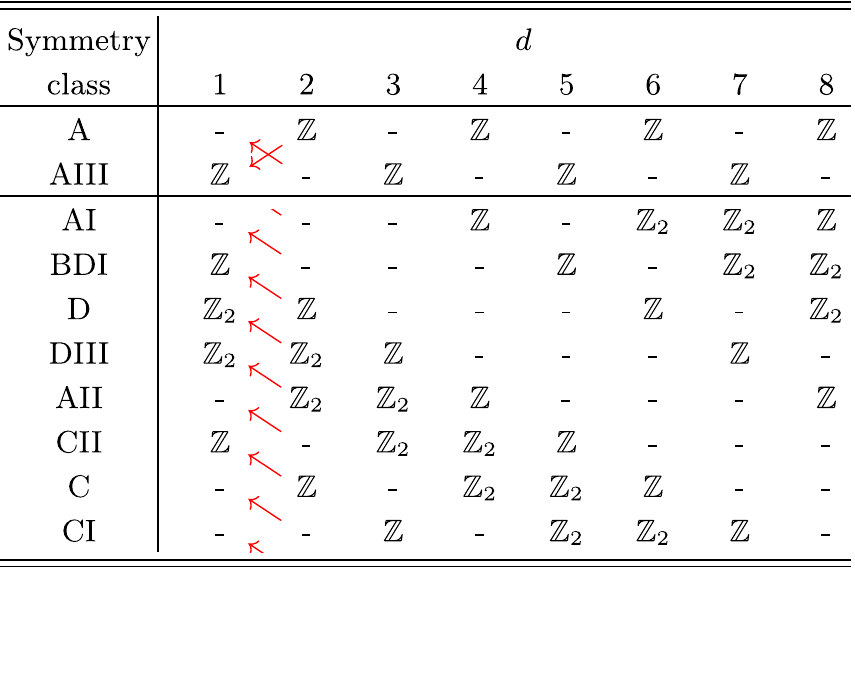}
\caption{%
  (Color online) Topological classification of the ten symmetry classes in
  different dimensions. 
  Combinations of symmetry class and dimensionality which support non-trivial topological
  invariants are indicated by the type of the topological invariant
($\mathbb{Z}$ or
  $\mathbb{Z}_2$). Classes which support only trivial insulators
  are denoted by `-'.  The arrows indicate the change of symmetry class
  upon dimensional reduction as discussed in the main text.  The topmost
  symmetries A and AIII (which do not have any anti-unitary symmetries) transform into
  each other, whereas the remaining 8 classes (with anti-unitary symmetries)
  exchange cyclically.  The dimensional
  reduction changes the symmetry class, but preserves the topological invariant
(`-',
  $\mathbb{Z}$, or $\mathbb{Z}_2$).
  }\label{tab:az}
\end{table}

We now turn to prove that for localized systems topological invariants
$\mathcal{Q}(H)$ and $\mathcal{Q}(H_{d-1})$ are identical. This
correspondence was proven in 1D in Ref.~\onlinecite{Ful11}, so here we
accomplish the proof in higher dimensions.

First of all, we observe that a topologically trivial Hamiltonian can be
deformed into a bunch of completely decoupled localized orbitals without
closing its gap.  In a sufficiently large system, this also means that the gap
of $H_{d-1}(\bm{k})$ does not close during this process.  For a system of
decoupled orbitals, $r(\bm{k})$ and accordingly $H_{d-1}(\bm{k})$ are
momentum-independent (and hence $H_{d-1}(\bm{k})$ is topologically trivial).
This means that a sufficiently large system with trivial $H(\bm{k})$ maps onto
a trivial $H_{d-1}(\bm{k})$ under the scheme of dimensional reduction outlined
above.

\begin{figure}[tb]
 \includegraphics[width = 0.3\textwidth]{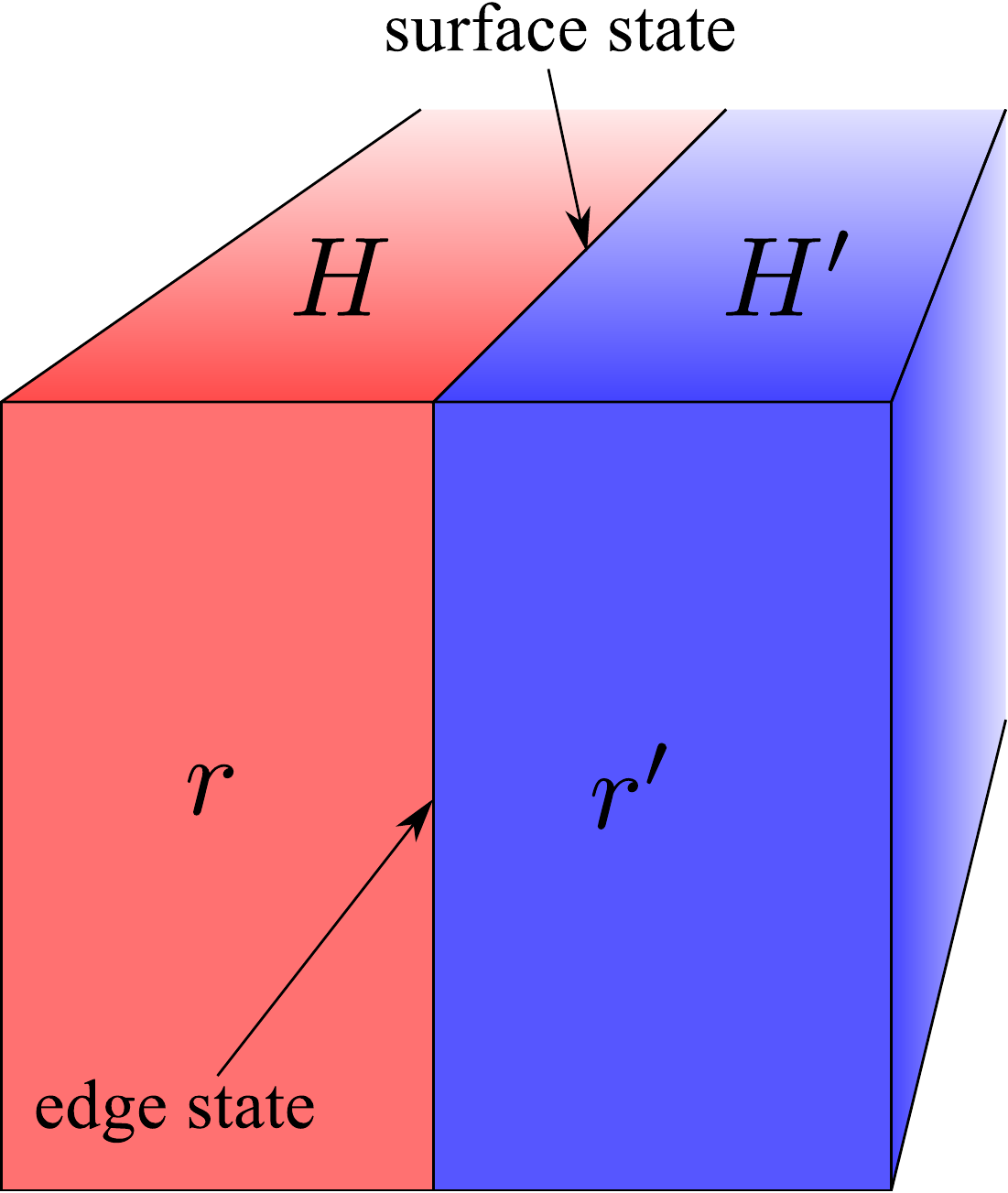}
 \caption{%
 A system in $d$ dimensions consisting out of two parts with
 different Hamiltonians $H$ and $H'$. Reflection blocks of the scattering
 matrix $r$ and $r'$ are used to define the lower dimensional Hamiltonians
 $H_{d-1}$ and $H_{d-1}'$. We prove the correspondence between topological
 invariants in $d$ and $d-1$ dimensions using the relation between the surface
 state at the interface between $H$ and $H'$ and the edge state at the
 interface between $H_{d-1}$ and $H_{d-1}'$.
 }\label{fig:interface}
\end{figure}

Let us now consider an interface between two systems with different bulk
Hamiltonians $H$ and $H'$, shown in Fig.~\ref{fig:interface}.  If the
Hamiltonians $H_{d-1}$ and $H_{d-1}'$ constructed out of reflection blocks of
the two systems have different topological invariants, a topologically
protected zero energy edge state in $d-1$ dimensions must appear at the
interface between them.  Recalling that a zero energy edge state in $d-1$
dimension corresponds to a perfectly transmitting mode of the original
$d$-dimensional system, we conclude that $H$ and $H'$ have different
topological invariants.

Conversely, if $H$ and $H'$ have different topological invariants, there
exists a transmitting mode at the interface between two parts of the system,
which appears irrespective of system size and microscopic details of the
interface.  This means that it is not possible to construct an interface
between $H_{d-1}$ and $H_{d-1}'$ which would be completely gapped.

Finally, the edge states in $d-1$ dimension have to have the same group
properties as the surface states in $d$ dimensions, leading us to the
conclusion that $\mathcal{Q}(H) = \mathcal{Q}(H_{d-1})$, as we set out to
prove.  The topology-preserving property of our dimensional reduction
procedure is the same as that of the mapping from a general $d$-dimensional
Hamiltonian to a $d+1$-dimensional Hamiltonian presented in
Ref.~\onlinecite{Teo10}.

At this point one might wonder why we apply the dimensional reduction only
once.  Indeed, the reduced Hamiltonian $H_{d-1}$ can be straightforwardly
approximated by a tight-binding Hamiltonian on a $d-1$ dimensional
lattice using a Fourier transform. This allows to repeat the procedure of
dimensional reduction until we arrive at a zero dimensional Hamiltonian. We
stop at the first dimensional reduction for practical purposes, since the
advantage of considering only Fermi level properties is achieved already
at the first step.

\section{Results for one--three dimensions}
\label{charge}

\subsection{Topological invariant in 1D}
\label{onedcharge}

We begin by verifying that we recover the 1D results of Ref.~\onlinecite{Ful11}, where the topological invariant was related to the scattering matrix without going through the procedure of dimensional reduction. Dimensional reduction in this case brings us to a zero-dimensional Hamiltonian. The topological invariant of a zero-dimensional
Hamiltonian without symmetry between positive and negative energies
(symmetry classes A, AI, and AII) is given just by the number of states
below the Fermi level. In class AII Kramers' degeneracy makes this
number always even. In addition, in 0D there exist two $\mathbb{Z}_2$
topological insulators in symmetry classes D and BDI. The topological
number is in that case the ground state fermion parity, or the Pfaffian
of the Hamiltonian in the basis where it is antisymmetric. To summarize,
\begin{subequations}
\begin{align}
 \mathcal{Q}(H) &= \nu(H), && \textrm{for A, AI, and AII,}\\
 \mathcal{Q}(H) &= \Pf\, iH, &&\textrm{for D and BDI,}
\end{align}
\end{subequations}
where $\nu(A)$ denotes the number of negative eigenvalues of the Hermitian matrix
$A$. Substituting $H$ from Eqs.~\eqref{eq:reducedham} yields
\begin{subequations}
\label{eq:1dcharge}
\begin{align}
 \mathcal{Q} &= \nu(r), &&\textrm{for AIII, BDI, and CII}\\
 \mathcal{Q} &= \Pf \, ir, &&\textrm{for DIII,}\\
 \mathcal{Q} &= \Pf 
\begin{pmatrix}
0 & i r \\
-i r^T & 0              
\end{pmatrix} \nonumber\\
&= \det r, &&\textrm{for D.}
\end{align}
\end{subequations}
We confirm that the Eqs.~\eqref{eq:1dcharge} are in agreement with
Ref.~\onlinecite{Ful11}.

\subsection{Topological invariant in 2D} \label{zeropolecharge} 

Starting from 2D, the dimensional reduction brings us to a 1D Hamiltonian. In this subsection we first review the known expressions for the topological invariants of 1D Hamiltonians, and then describe how to efficiently evaluate it
for the effective Hamiltonian \eqref{eq:reducedham}. The $\mathbb{Z}$
topological insulators in 1D (classes AIII, BDI, and CII) are characterized
by a winding number\cite{Zak89,Ryu02}
\begin{align}
\label{eq:chiralh}
 H(k) &\equiv \begin{pmatrix}
         0 & h(k) \\
         h^\dagger (k) & 0
        \end{pmatrix},\\
        \mathcal{Q}(H) 
        &= \frac{1}{2\pi i} \int_0^{2\pi} \!\! dk  \frac{d}{dk} \log
        \det h(k), \nonumber\\
        &\hspace{0.3\linewidth}\text{for AIII, BDI, and CII}.
\label{eq:windingh}
\end{align}
The topological invariant for the Hamiltonian in class D is given
by Kitaev's formula\cite{Kit01}
\begin{equation}
 \mathcal{Q}(H) = \sign\left[\frac{ \Pf \, H(0) }{\Pf \, H(\pi)}\right],
 \quad\text{for D}.
\end{equation}
Finally, in class DIII the expression for the topological invariant was derived
in Ref.~\onlinecite{Qi10}:
\begin{align}
\label{eq:1dd3}
 \mathcal{Q}(H) 
 &= \frac{\Pf[U_{\cal T} h(\pi)]}{\Pf[U_{\cal T} h(0)]}
 \exp \Biggl[ -\tfrac12\int_0^\pi \!\!dk\,\frac{d}{dk}\log \det h(k)
  \Biggr] \nonumber\\
&=\frac{\Pf[U_{\cal T} h(\pi)]}{\Pf[U_{\cal T} h(0)]}
\frac{\sqrt{\det h(0)}}{\sqrt{\det h(\pi)}}, \quad\text{for DIII,}
\end{align}
where the square root is defined through analytic continuation over the first
half of the Brillouin zone, $h$ is defined by Eq.~\eqref{eq:chiralh}, and $U_{\cal T}$ is the unitary part of the time reversal operator ${\cal T} = U_{\cal T} {\cal K}$.

Substituting Eq.~\eqref{eq:reducedham} into the expressions for topological
charge we get
\begin{subequations}
\begin{align}
 \mathcal{Q} &= \frac{1}{2\pi i} \int_0^{2\pi} \!\! dk  \frac{d}{dk} \log
        \det r(k), &&\textrm{for A, C, D} \label{eq:windingr}\\
 \mathcal{Q} &= \frac{\Pf[U_{\cal T} r(\pi)]}{\Pf[U_{\cal T} r(0)]}
\frac{\sqrt{\det r(0)}}{\sqrt{\det r(\pi)}}, &&\textrm{for AII,} \label{eq:2da2}\\
 \mathcal{Q} &= \sign\left[\frac{ \Pf \, r(0) }{\Pf \, r(\pi)}\right], &&\textrm{for DIII.}
\end{align}
\end{subequations}

In order to efficiently evaluate the integral given in
Eq.~\eqref{eq:windingh}, and the analytic continuation in Eq.~\eqref{eq:2da2} using Eq.~\eqref{eq:rdef}, we define
a new variable $z = e^{i k}$. Then we perform an analytic continuation of
$\det \, r(z)$ to the complex plane from the unit circle $|z| = 1$. To find zeros and poles of $\det \, r(z)$ we use
\begin{equation}\label{eq:numdenom}
   \det r(z) = 
  \det 
\begin{pmatrix}
A- Z_1(k) & B \\
C & D
\end{pmatrix} \bigg/ \det 
   \left[ A -Z_{1}(k) \right],
\end{equation}
where
\[
Z_1(k) = \begin{pmatrix}  0 & e^{ik} \\ e^{-ik} & 0 \end{pmatrix}
  = \begin{pmatrix}  0 & z \\ z^{-1} & 0 \end{pmatrix};
\]
Equation \eqref{eq:numdenom} follows from Eq.~\eqref{eq:rdef} and the
determinant identity
\begin{equation}
\det(D - C M^{-1}B) =
 \det \begin{pmatrix}
       M & B \\
       C & D
      \end{pmatrix} \bigg/ \det M .
\end{equation}

Since both the numerator and the denominator of Eq.~\eqref{eq:numdenom}
are finite at any finite value of $z$, the roots of the numerator $z_n$
are the zeros of $\det r(z)$, and the roots of the denominator $w_n$ are
the poles. In App.~\ref{app:poles} we show that due to unitarity of the
scattering matrix, the poles of $\det r(z)$ never cross the unit circle. By
multiplying the second column of the numerator of Eq.~\eqref{eq:numdenom}
by $z$ we bring the problem of finding roots $z_n$ of this numerator to
the generalized eigenvalue problem,
\begin{equation}\label{eq:generalizedA}
  \begin{pmatrix}
    S_{1,1} & -1 & S_{1,2} \\
    S_{\bar1,1} & 0 & S_{\bar1,2} \\
    S_{2,1} & 0 & S_{2,2}
  \end{pmatrix} \psi_n =
  z_n \begin{pmatrix}
    0 & -S_{1,\bar1} & 0 \\
    1 & -S_{\bar1,\bar1} & 0 \\
    0 & -S_{2,\bar1} & 0
  \end{pmatrix} \psi_n,
\end{equation}
which can be efficiently evaluated. The roots $w_n$ of the denominator
can also be found by solving the generalized eigenvalue problem,
\begin{equation}\label{eq:generalizedpoles}
  \begin{pmatrix}
    S_{1,1} & -1 \\
    S_{\bar1,1} & 0
  \end{pmatrix} \psi_n =
  w_n \begin{pmatrix}
    0 & -S_{1,\bar1}  \\
    1 & -S_{\bar1,\bar1}
  \end{pmatrix} \psi_n.
\end{equation}

Since the poles of $\det r(z)$ never cross the unit circle, in classes A,
C, and D the topological invariant is given by
\begin{equation}\label{eq:top_a}
  \mathcal{Q} = \#\{z_n: |z_n| <1 \} - N_1, \quad 
  \text{for A, C, and D},
\end{equation}
i.e., the number of $z_n$'s inside the unit circle minus the number of modes
$N_1$ in the direction 1.  In class AII (quantum spin Hall insulator)
the topological invariant is given by
\begin{equation}\label{eq:top_aii}
  \mathcal{Q} = 
  \frac{\prod_n \sqrt{i\frac{ 1+ z_n}{1-z_n}}}{\prod_n \sqrt{(-i)\frac{ 1+
  w_n}{1-w_n}}}
  \times \frac{\Pf \, U_{\cal T} r(\pi)}{  \Pf \, U_{\cal T} r(0)}
  \quad\text{for AII},
\end{equation}
with the branch cut of the square root along the negative real axis. Note that
the linear fractional transformation $z\mapsto i(1+z) / (1-z)$ maps the upper
half of the unit circle onto the negative real axis. In symmetry class DIII
the evaluation of the topological invariant is most straightforward, and yields
\begin{equation}\label{eq:top_diii}
 \mathcal{Q} = \sign\left[\frac{ \Pf \, r(0) }{\Pf \, r(\pi)}\right]
 \quad\text{for DIII}.
\end{equation}

The physical meaning of the topological invariant in class A is quantized
pumping of charge as a response to magnetic flux. In the quantum spin
Hall insulator in class AII the invariant can be interpreted either
as time-reversal polarization pumping\cite{Fu06}, or as pumping of spin
 which is quantized along an unknown axis.\cite{Gda11, GdaXX} 
In the superconducting classes C, D, and DIII it is an analogous thermal 
or gravitational response.\cite{Ryu10-2, Wan10}

\subsection{Topological invariant in 3D}
\label{z2charge}

Turning now to 3D, we need to consider topological invariants of 2D
Hamiltonians.  The symmetry class with the simplest expression for the
topological invariant in terms of the scattering matrix in 3D is AII.
The 2D topological invariant of a system in class DIII (into which AII
transforms upon dimensional reduction) is a product\cite{Qi10} of the
topological invariants \eqref{eq:1dd3} of 1D Hamiltonians obtained by
setting one of the momenta to $0$ or $\pi$,
\begin{equation}\label{eq:2d_diii}
  \mathcal{Q}[H(k_1, k_2)] = \mathcal{Q}[H(k_1,0)]\mathcal{Q}[H(k_1,\pi)],
\end{equation}
with $\mathcal{Q}[H(k_1)]$ given by Eq.~\eqref{eq:1dd3}.
Substituting Eq.~\eqref{eq:reducedham} into this expression we obtain
\begin{multline}\label{eq:top_aii_3d}
\mathcal{Q} = \frac{\Pf[U_{\cal T} r(\pi,0)]}{\Pf[U_{\cal T} r(0,0)]}
\frac{\sqrt{\det r(0,0)}}{\sqrt{\det r(\pi,0)}} \times \\
\frac{\Pf[U_{\cal T} r(\pi,\pi)]}{\Pf[U_{\cal T} r(0,\pi)]}
\frac{\sqrt{\det r(0, \pi)}}{\sqrt{\det r(\pi, \pi)}}\textrm{,\quad for AII.}
\end{multline}

Direct evaluation of the Hamiltonian topological invariant in 2D in classes
with nontrivial Chern number (A, C, D), and in class AII is hard because of
the need to fix the gauge throughout the Brillouin zone\cite{Tho82, Fu06}. It
is usually more efficient to use a method which relies on the real space
structure of $H$ evaluated in a single point in momentum space.\cite{Lor10,
Rin11, Pro11, Lor11} These methods using the Bott index or a similar expression
for the topological invariant require the so-called band-projected position
operators: $x_P = P \exp(2\pi i x) P$ and $y_P = P \exp(2\pi i y) P$. Here
$P$ is the projector on the states of the Hamiltonian with negative
energies, and $x$ and $y$ are the coordinate operators in the unit cell
of the system. In order to evaluate these operators in our case we note
that the eigenvalues of the effective Hamiltonian in the symmetry classes
of interest approach $\pm 1$ when the original system becomes localized. In
that case $P = (1 - r)/2$ [with $r \equiv r(0,0)$], and we can avoid the need
to calculate the projector explicitly if we approximate $x_P$ and $y_P$ by
\begin{gather}
 x_P \approx (1+r)/2 + (1-r) e^{2\pi i x} (1-r) / 4,\\
 y_P \approx (1+r)/2 + (1-r) e^{2\pi i y} (1-r) / 4.
\end{gather}
Using the 2D Hamiltonian expressions from Ref.~\onlinecite{Has10} we arrive
at a scattering formula for the 3D topological invariant,
\begin{equation}\label{eq:bott}
 \mathcal{Q} = \frac{1}{2\pi}\imag \tr \log[x_P y_P x_P^\dagger y_P^\dagger]\textrm{,\quad for AIII, CI, DIII.}
\end{equation}

The symmetry class CII in 3D transforms upon dimensional reduction to
class AII in 2D. The expressions for the Pfaffian-Bott index required to
calculate the topological invariant for a 2D Hamiltonian in class AII are
quite involved. We do not give them here, but refer the interested reader
to Eqs. (7), (9), and (10) of Ref.~\onlinecite{Lor10}.

\subsection{Weak invariants}

All of the algorithms described above apply directly to the weak topological
invariants.\cite{Fu07, Moo07, Qi08} In order to evaluate a weak invariant one just
needs to eliminate one of the dimensions by setting the momentum along that
dimension to either $0$ or $\pi$, and to evaluate the appropriate topological
invariant for the resulting lower dimensional system. The only caveat is that
since weak topological indices do not survive doubling of the unit cell,
the thickness of the system in the transverse direction should be equal
to the minimal unit cell. In the same fashion (eliminating one momentum
or more) one can calculate the presence of surface states\cite{Sch10}
in chiral superconductors and Fermi arcs\cite{Wan11} in 3D systems.

\section{Applications and performance}
\label{numerics}

\subsection{Performance}

The complexity of the Hamiltonian expressions scales with linear system
size $L$ as $L^2$ in 1D, and as $L^{3d}$ in higher dimensions. In contrast,
the complexity of the scattering matrix expressions scales proportionally to
$L$ in 1D and to $L^{3d-3}$ in higher dimensions.\cite{Ala73, Li08}
All the subsequent operations have the same or a more favorable scaling. We
use the algorithm of Ref.~\onlinecite{Wim11} to calculate the Pfaffian of an
arbitrary skew-symmetric matrix.

We have verified that using the scattering matrix method allows to
efficiently calculate the topological invariant of a quantum Hall system
and of the BHZ model\cite{Ber06} discretized on a square lattice with a
size of $1000\times 1000$. This improves considerably on
previously reported\cite{Lor10, Pro11a} results of up to $50
\times 50$ lattice sites for the BHZ model.

In 3D the improvement in performance is not as large because the values of
$L$ that we can reach are smaller.  Nevertheless, we have confirmed that it is
possible to calculate the topological invariant of 3D systems in classes AII
and DIII using a 4-band model on a cubic lattice with system size $50\times
50\times 50$. This is a significant improvement over the $12\times 12\times
12$ size, reported for Hamiltonian-based methods.\cite{Has11}

In addition to tight-binding models, our method applies very naturally to
various network models,\cite{Cha88, Cho97, Ryu10-1} which are favorite
models for the phase transitions. Hamiltonian-based approaches are not
applicable to the network models, since those only have a scattering matrix,
and no lattice Hamiltonian. We have checked that calculating a topological
invariant of the Chalker-Coddington network model of size $1000\times 1000$
only takes several minutes on modern hardware.

\subsection{Finite size effects}

\begin{figure}[tb]
 \centerline{\includegraphics[width=.7\linewidth]{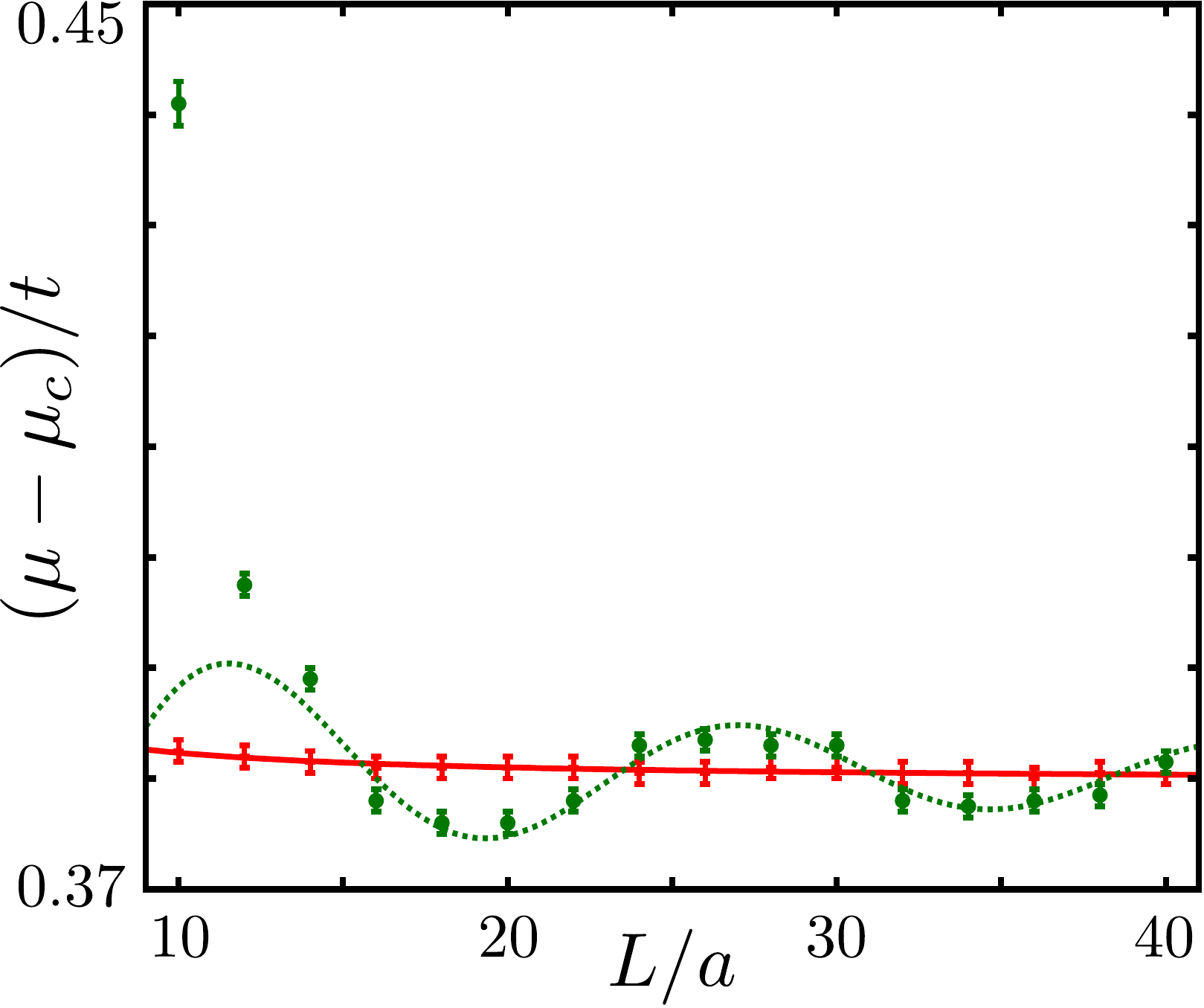}}
\caption{%
The value of the chemical potential $\mu_c$ where the ensemble averaged
topological invariant equals to 0.5, as a function of system size $L$. Red: 
topological invariant defined in terms of the scattering matrix,
from Eq.~\eqref{eq:top_a}. Green: topological invariant obtained from
the Hamiltonian expression of Ref.~\onlinecite{Lor10}. Lines represent fits 
as described in the text.
}\label{fig:finitesize} 
\end{figure}
The expressions for the topological invariant given in terms of the
scattering matrix in Sec.~\ref{charge} do not coincide with $\mathcal{Q}(H)$
very close to the transition. This is a finite size effect.  In order to
estimate the importance of finite size effects we have computed the
shift of the transition point between the $n=0$ and $n=1$ plateaus of
a disordered quantum Hall system as a function of size. We have used a
square lattice discretization (lattice constant $a$) of a single band tight binding model with nearest neighbor hopping $t=1$. The magnetic flux per unit cell of the lattice was fixed at $0.4\hbar/e$. We used on-site disorder homogeneously distributed on an interval $[-0.05, 0.05]$.

The transition point is defined as the value of the chemical potential $\mu_c$
at which the disorder-averaged topological invariant equals 0.5.  We have
compared two expressions for the topological invariant: the scattering matrix
expression \eqref{eq:top_a} and the Hamiltonian expression from
Ref.~\onlinecite{Lor10}.  The results are shown in Fig.~\ref{fig:finitesize}.
We fit the data obtained via the scattering matrix approach to the function
$f(L) = c_1 + c_2/L$ obtaining a value $c_2 \approx 0.026$.  In the case of
the expression of Ref.~\onlinecite{Lor10}, the finite size effect are best fit
to the function $g(L) = c_1' +c_2' \sin (c_3 L + c_4) / L$, with $c_2' \approx
0.116$.  We conclude that the finite size effects of our algorithm are
significantly lower.

\subsection{Applications}

\begin{figure*}[tb]
  \centering
  \includegraphics[width=.8\textwidth]{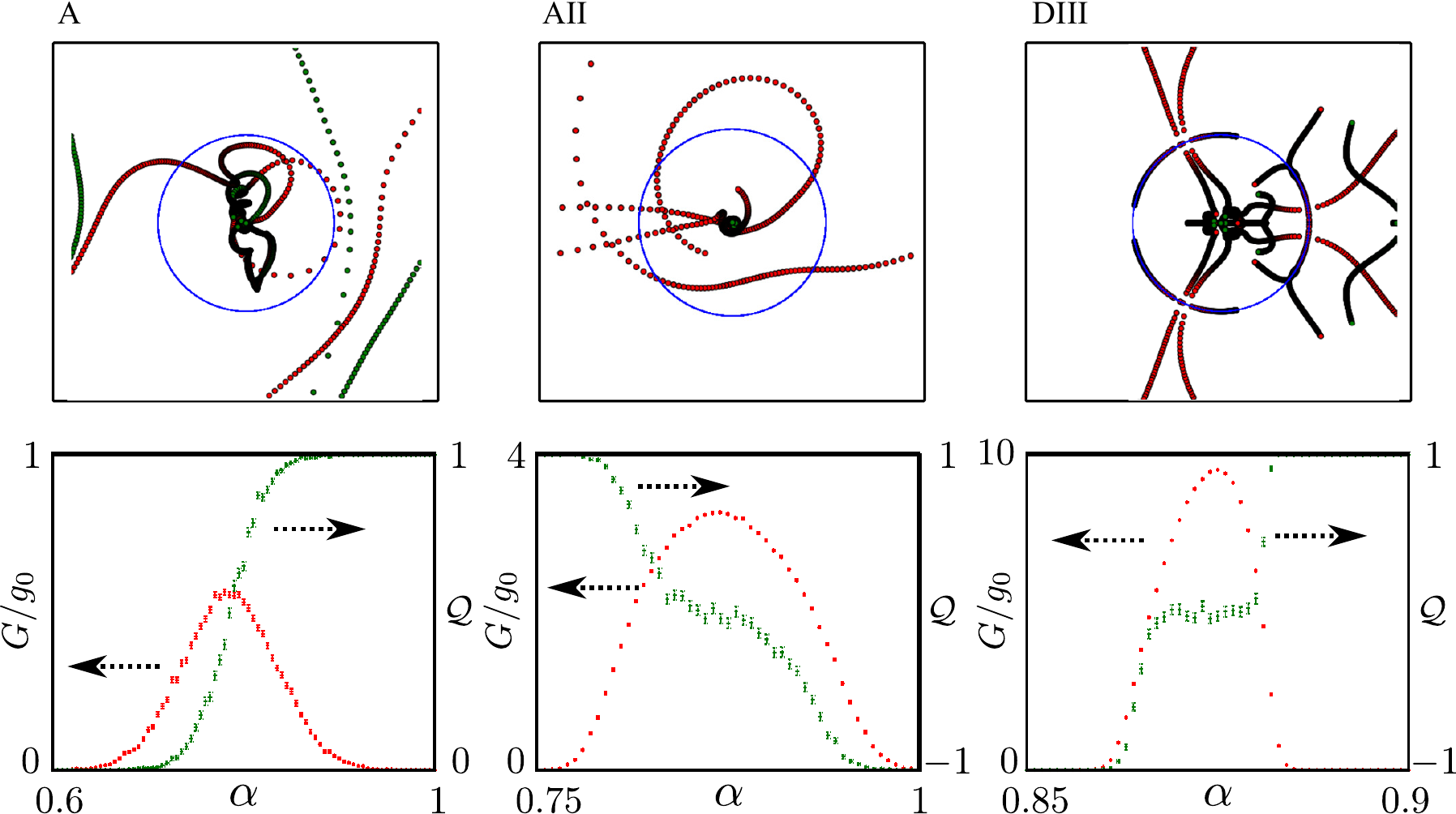}
  \caption{%
  Top panel: Evolution of the poles (green dots) and the zeros (red dots)
  of $\det r(z)$ as a function of a parameter $\alpha$ which tunes through
  the topological phase transition in classes A, AII, and DIII in 2D. Shown
  is the complex plane with the unit circle $|z|=1$ indicated in blue.
  Time-reversal symmetry in AII and DIII implies that for every zero/pole at
  $z_0$ there is additionally one at $1/z_0$. In DIII, there is additional
  particle-hole symmetry which additionally dictates zeros/poles at $z_0^*$
  and $1/z_0^*$.The phase transition happens when at least one of the
  zeros crosses the unit circle.  This event coincides with a change of
  the topological invariant $\mathcal{Q}$ (green) defined by Eqs.~(\ref{eq:top_a} -- \ref{eq:top_diii})}, as shown in the bottom panels.
  \label{fig:fingerprint}
\end{figure*}
In 2D we illustrate our approach by applying it to network models in
classes A, AII, and DIII. In class A we use the Chalker-Coddington network
model.\cite{Cha88} In classes AII and DIII we have used the quantum spin
Hall network model of Ref.~\onlinecite{Ryu10-1}. In class DIII we have
set the link phases to zero in order to ensure particle-hole symmetry. In each of these cases the parameter which tunes through the transition is the angle $\alpha$ related to reflection probability at a node of the network by $R = \cos^2 \alpha$.

Our results are summarized in Fig.~\ref{fig:fingerprint}.  Top panels show the
evolution of zeros and poles of $\det r(z)$ across the phase transition ---
the `fingerprint' of a topological phase transition.\cite{movies} There are no
symmetry constraints on this fingerprint in class A.  The time-reversal
symmetry ensures that for every zero or pole at $z_0$ there is another one at
$1/z_0$.  The particle-hole symmetry translates into the mirror symmetry with
respect to the real axis: for every zero or pole at $z_0$ there is one at
$z_0^*$.  The bottom panels show the behavior of the topological invariant and
of the conductance $G = \tr t^\dagger t$, with $t$ the
transmission matrix through the system.  The simulations were performed on
systems of size $300\times 300$ in each of the symmetry classes and averaged
over 1000 samples. The presence of plateaus around zero in the curves for the
topological invariant coincides with the presence of a metallic phase in the
phase diagram of symmetry classes AII and DIII.

\begin{figure}[tb]
 \includegraphics[width=.8\linewidth]{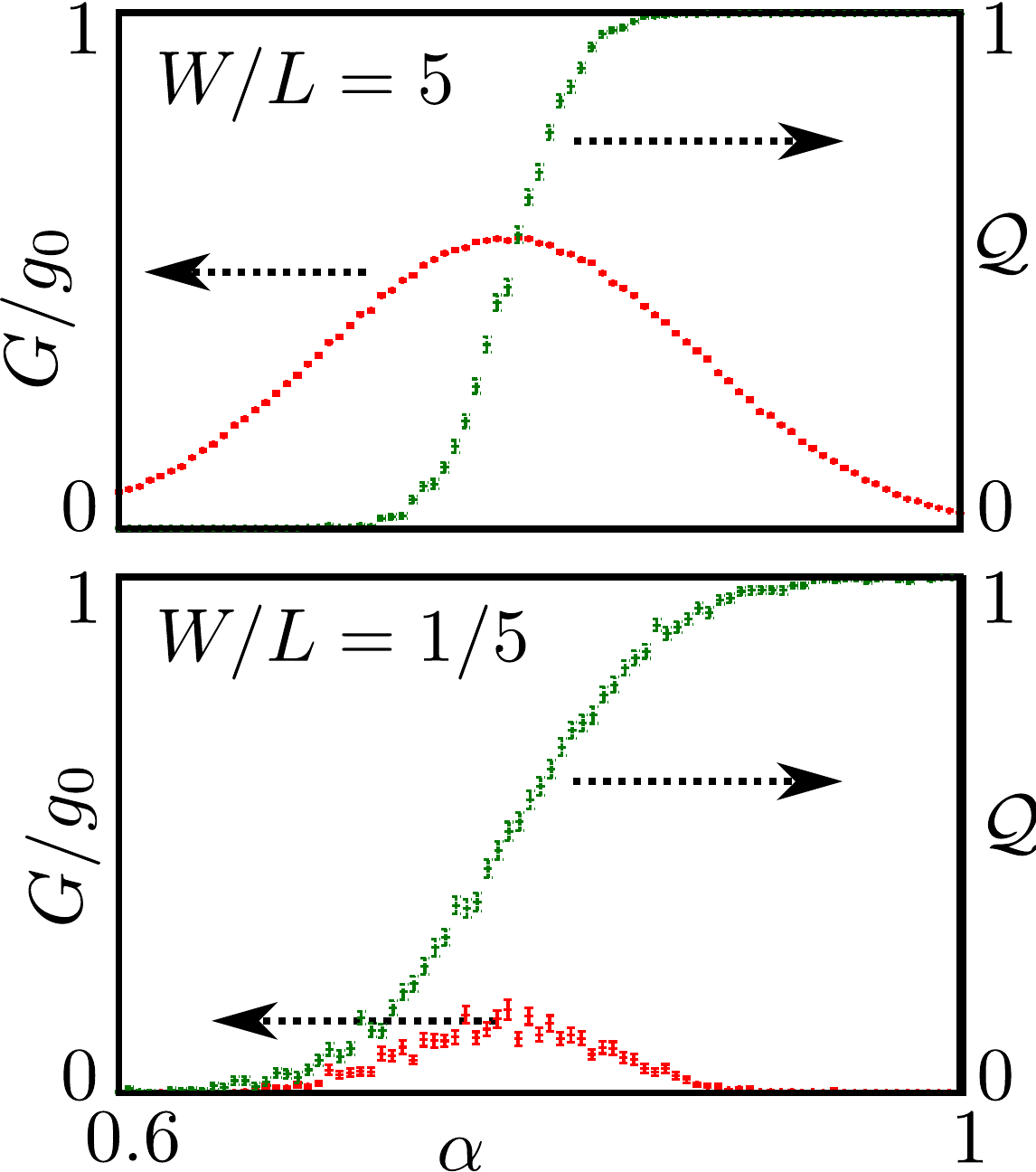}
\caption{%
Average topological invariant $\mathcal{Q}$ \eqref{eq:top_a} and longitudinal
conductance $G$ of a disordered
quantum Hall sample for different aspect ratios as a function of the
mixing angle $\alpha$.
 }\label{fig:aspectratio} 
\end{figure}

Although we introduced the topological invariant through transport properties,
it does not always have the same features as the conductance. The topological
invariant characterizes winding of scattering modes in the transverse
direction. Accordingly, in a system with a large ratio of width $W$ to the
length $L$, the width of the transition of the topological invariant is
reduced. The width of the peak in the conductance, on the contrary, is
reduced if $W/L$ becomes small. This is in agreement with what we observe
in numerical simulations. We have calculated the topological
invariant and conductance averaged over 1000 disorder realizations
in the Chalker-Coddington network model in systems with $W=300$ and $L =
60$ and vice versa. The results are shown in Fig.~\ref{fig:aspectratio}
and they agree with our expectations.

\begin{figure}[tb]
 \includegraphics[width=.7\linewidth]{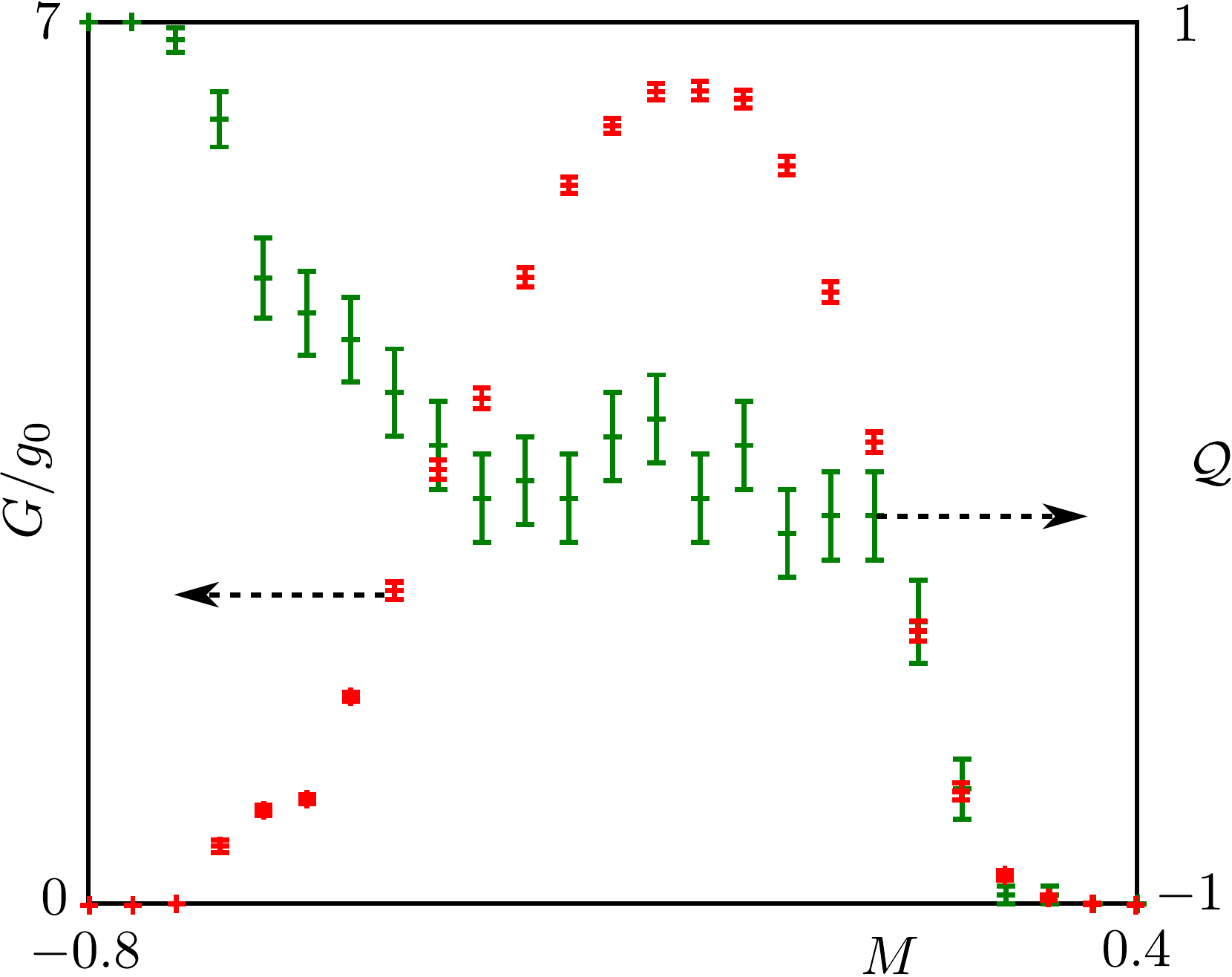}
\caption{%
Conductance and topological invariant \eqref{eq:top_aii_3d} for a
disordered 3D topological
insulator in class AII.
}\label{fig:3dconductionindex}
\end{figure}

We have also studied a 3D topological system in class AII on a
cubic lattice. We have used a simplified version of the Hamiltonian of Ref.~\onlinecite{Zha09}:
\begin{equation}
H =
\begin{pmatrix}
 {\cal M}({\bf k}) & v k_z            & 0                 & v k_-  \\
 v k_z           & -{\cal M}({\bf k}) & v k_-           & 0        \\
 0                 & v k_+            & {\cal M}({\bf k}) & -v k_z \\
 v k_+           & 0                  & -v k_z          & -{\cal M}({\bf k})
\end{pmatrix}
- \mu
\end{equation}
discretized on cubic lattice with lattice constant $a$, where $k_\pm = k_x \pm ik_y$, and 
${\cal M}({\bf k}) = M - \alpha k^2$. The Hamiltonian parameters 
were chosen to be $\alpha = a^2$, $v = a$. We chose $\mu = \mu_0 + \delta \mu$ with $\mu_0 = 0.4$, and $\delta\mu$ being a random uncorrelated variable uniformly distributed in the interval $[-2, 2]$. The topological invariant defined by
Eq.~\eqref{eq:top_aii_3d} as well as the longitudinal conductance for a
$20\times 20\times 20$ system averaged over 100 disorder realizations are
shown in Fig.~\ref{fig:3dconductionindex} as a function of $M$. We observe that, 
analogously to the two-dimensional case, the presence of a metallic phase is 
accompanied by a plateau in the topological charge at a value of zero.

\section{Conclusion}
\label{conclusion}

In conclusion, we have introduced a procedure of dimensional reduction which
relates a scattering matrix of a $d$-dimensional system to a Hamiltonian
in $d-1$ dimensions with a different symmetry class, but with the same
topological invariant as the original system. When applied repeatedly
this dimensional reduction procedure serves as an alternative derivation
of the Bott periodicity of topological insulators and superconductors.

Since our approach uses only Fermi surface properties it is much more
efficient than existing alternatives which require the analysis of the
full spectrum. We have described how to implement our method efficiently
in all the symmetry classes in 1--3 dimensions. We have verified that it
allows to analyze much larger systems than previously possible.

This paper focused on the description of the method and we only touched
on a few applications at the end. More applications can be envisaged and
we believe that the scattering approach will lead to the discovery of new
observable physics at topological phase transitions.

\acknowledgments

This research was supported by the Dutch Science Foundation NWO/FOM. We
thank B.~B\'{e}ri, L.~Fu, and J.~Tworzyd{\l}o for useful discussions. We are
especially grateful to M.~Wimmer for explaining the efficient method to
calculate transport properties and to C.~W.~J.~Beenakker for expert advice.

\appendix

\section{Introduction to discrete symmetries}
\label{app:symmetries}

Here we define the three core discrete symmetries, and the corresponding symmetry constraints on the Hamiltonians and on the scattering matrices. We also specify how to choose the symmetry representation we used in Fig.~\ref{fig:symmetries}.

\subsection*{Definitions and properties of discrete symmetries}

The discrete symmetries are defined as follows: The time
reversal symmetry operator $\mathcal{T}$ is an anti-unitary operator. When it
is applied to an arbitrary eigenstate $\psi$ of the Hamiltonian $H$ at
energy $\varepsilon$, returns an eigenstate of the Hamiltonian at the same energy:
\begin{subequations}\label{eq:symmetries}
\begin{equation}
 H\psi = \varepsilon\psi \Rightarrow H \mathcal{T}\psi = \varepsilon \mathcal{T}\psi
\end{equation}
On the other hand, the anti-unitary particle-hole symmetry
operator
$\mathcal{P}$ returns
an eigenstate with opposite energy when applied to any eigenstate of the Hamiltonian:
\begin{equation}
 H\psi = \varepsilon\psi \Rightarrow H \mathcal{P}\psi = -\varepsilon \mathcal{P}\psi
\end{equation}
Chiral symmetry $\mathcal{C}$ also reverses energy, but unlike
the other two has a unitary operator.
\end{subequations}

All three symmetries $\mathcal{T}, \mathcal{P}, \mathcal{C}$ are $\mathbb{Z}_2$ symmetries, so the symmetry operators must square to a phase factor.

In an arbitrary basis the symmetry operators are represented by
\begin{equation}
  {\cal T} = U^\pdag_{\cal T}{\cal K}, \quad
  {\cal P} = U^\pdag_{\cal P}{\cal K}, \quad
  {\cal C} = U^\pdag_{\cal C},
\end{equation}
with $\mathcal{K}$ denoting the complex conjugation, and $U_\mathcal{T}$, $U_\mathcal{P}$, and $U_\mathcal{C}$ unitary matrices. 
Since $\exp(i\phi) = \mathcal{T}^2 =
U_{\mathcal{T}} U_{\mathcal{T}}^*$, and its determinant $\det
U_{\mathcal{T}} U_{\mathcal{T}}^*$ is real, we obtain
$U_{\mathcal{T}} U_{\mathcal{T}}^* = \pm 1$; similarly $U_{\mathcal{P}}
U_{\mathcal{P}}^* = \pm 1$. In other words, every anti-unitary
symmetry comes in two flavors, squaring either to $+1$ or to $-1$. No such constraint applies to chiral symmetry, which may square to an arbitrary phase factor $\exp(i \phi)$. This factor however can always be set to zero by choosing $U_\mathcal{C} \rightarrow U_\mathcal{C} \exp(-i\phi/2)$.

The symmetry constraints on the Hamiltonian 
\begin{subequations} 
\begin{align}
 H &= {\cal T} H {\cal T}^{-1} = U_{\cal T}^\pdag H^* U^\dag_{\cal T} \label{eq:trevsym}\\
 H &= -{\cal P} H {\cal P}^{-1} = -U_{\cal P}^\pdag H^* U^\dag_{\cal P} \label{eq:pholesym}\\
 H &= -{\cal C} H {\cal C}^{-1} =- U^\pdag_{\cal C} H U_{\cal C}^\dag\label{eq:chiralsym}.
\end{align} 
\end{subequations}
follow immediately from the definition of the symmetries, Eq.~\eqref{eq:symmetries}.

\subsection*{Relation between discrete symmetries and translational invariance}

In addition to the basic properties, the discrete symmetries in periodic systems are required to commute with the coordinate operator. 
So for any Bloch wave written as
\begin{equation}
\psi(\bm{r}) = e^{i\bm{p} \bm{r}} \psi(0),
\end{equation}
with $\bm{r}$ coordinate in a translationally-invariant system, and $\psi(0)$ the wave function inside a single unit cell,
the action of the symmetry operators is:
\begin{subequations}
 \begin{align}
   \mathcal{T}\psi(\bm{r}) &= e^{-i\bm{p} \bm{r}} \mathcal{T}\psi(0) ,\\
   \mathcal{P}\psi(\bm{r}) &= e^{-i\bm{p} \bm{r}} \mathcal{P}\psi(0) ,\\
   \mathcal{C}\psi(\bm{r}) &= e^{i\bm{p} \bm{r}} \mathcal{C}\psi(0) .
 \end{align}
\end{subequations}
Since the velocity of a Hamiltonian eigenstate at energy $\varepsilon$ and
momentum $\bm{p}$ is $v=d\varepsilon/d\bm{p}$, time-reversal and chiral
symmetries reverse the velocity, while particle-hole symmetry
keeps the velocity invariant.

\subsection*{Symmetry constraints on scattering matrix}

In order to figure out what the symmetry constraints on the scattering
matrices are, we first review the basic properties of the
scattering matrices.
Scattering matrices act in the space of asymptotic scattering states outside
of the scattering region. This space contains two non-intersecting
subspaces: the subspace of incoming modes and the subspace of outgoing modes.
The incoming modes are all the plane waves with velocity in the direction of
the scattering region, and the outgoing modes are all the plane waves with
velocity pointing away from the scattering region. Let $\psi_n^{\text{in}}$ be
a basis in the space of incoming modes, and $\psi_n^{\text{out}}$ a basis in
the space of outgoing modes. Conventionally all the modes are normalized such
that current operator in the basis of $\psi^\text{in}$ is the identity matrix,
and the negative identity matrix in the basis of $\psi^\text{out}$.

The matrix elements of the scattering matrix $S$ satisfy
\begin{equation}
\label{eq:scattering_problem_app}
 (H - \varepsilon)\left(\psi_n^{\text{in}} + \sum_m S_{mn} \psi_m^{\text{out}} + \psi^{\text{loc}}\right) = 0,
\end{equation}
with $\psi^{\text{loc}}$ a wave-function localized near the scattering region.

As derived in the previous subsection, time-reversal and chiral symmetries
change the velocity to its opposite, while particle-hole symmetry leaves the velocity invariant. This means that scattering states transform under the discrete symmetries in the following manner:
\begin{equation}
\begin{aligned}
& \mathcal{T} \psi_n^{\text{in}} &=& \sum_m \left(V_\mathcal{T}\right)_{nm}
\psi_m^{\text{out}},\\
& \mathcal{T} \psi_n^{\text{out}} &=& \sum_m \left(Q_\mathcal{T}\right)_{nm}
\psi_m^{\text{in}},\\
& \mathcal{C} \psi_n^{\text{in}} &=& \sum_m \left(V_\mathcal{C}\right)_{nm}
\psi_m^{\text{out}},\\
& \mathcal{C} \psi_n^{\text{out}} &=& \sum_m \left(Q_\mathcal{C}\right)_{nm}
\psi_m^{\text{in}},\\
& \mathcal{P} \psi_n^{\text{out}} &=& \sum_m \left(V_\mathcal{P}\right)_{nm}
\psi_m^{\text{out}},\\
& \mathcal{P} \psi_n^{\text{in}} &=& \sum_m \left(Q_\mathcal{P}\right)_{nm}
\psi_m^{\text{in}}.
\end{aligned}\label{eq:mode_transformation}
\end{equation}
The additional constraints on the type of time-reversal, particle-hole, and chiral symmetries require
\begin{subequations}
\label{eq:vq_sign}
\begin{eqnarray}
 \pm 1 &=& \mathcal{T}^2 = V_\mathcal{T} Q_\mathcal{T}^*, \\
 \pm 1 &=& \mathcal{P}^2 = V_\mathcal{P}V_\mathcal{P}^* = Q_\mathcal{P}Q_\mathcal{P}^*, \\
 1 &=& \mathcal{C}^2 = V_\mathcal{C}Q_\mathcal{C}.
\end{eqnarray}
\end{subequations}

Applying time-reversal symmetry to Eq.~\eqref{eq:scattering_problem_app} and
using Eqs.~\eqref{eq:mode_transformation} we get
\begin{equation}
(H - \varepsilon)\left(V_\mathcal{T}\psi_n^{\text{out}} + \sum_m S_{mn}^* Q_\mathcal{T} \psi_m^{\text{in}}
 + \mathcal{T}\psi^{\text{loc}}\right) = 0,
\end{equation}
where we have also used that $H$ is time-reversal invariant. Comparing with
Eq.~\eqref{eq:scattering_problem_app}, we get
\begin{equation}
 S^{-1} = Q_\mathcal{T}^{T} S^* V_\mathcal{T},
\end{equation}
which we can be reduced to
\begin{subequations}
\label{eq:s_constraint}
\begin{equation}
 S = V_\mathcal{T}^T S^T Q_\mathcal{T}^*.
\end{equation}
Similarly, the chiral and the particle-hole symmetry constraints on $S$ are:
\begin{align}
 S &= V_\mathcal{C}^T S^\dagger V_\mathcal{C}^T,\\
 S &= V_\mathcal{P}^T S^* Q_\mathcal{P}^*.
\end{align}
\end{subequations}
Naturally, the constraints imposed by chiral and particle-hole symmetry
only hold at zero excitation energy, since these symmetries anti-commute
with the Hamiltonian. Finally, the symmetry constraints on the reflection
matrix are identical to Eqs.~\eqref{eq:s_constraint}, since $r$ is a diagonal
sub-block of $S$.

\subsection*{Choice of symmetry representation and mapping from scattering matrix to Hamiltonian symmetries}
The choice of symmetry representation is fully specified by choice of unitary
matrices $U_\mathcal{O}$, $V_\mathcal{O}$, and $Q_\mathcal{O}$ ($\mathcal{O} =
\mathcal{T}, \mathcal{P}, \text{ or }\mathcal{C}$). The symmetry representations
used in the main text were chosen to make the mapping from the reflection matrix
to an effective Hamiltonian most straightforward. In order to reach this aim, we
always choose $V_\mathcal{O}=\pm Q_\mathcal{O}$ for each of the three
symmetries. Whenever chiral symmetry is present, we use
\begin{equation} 
U_\mathcal{C} = \tau_z,\quad V_\mathcal{C} = Q_\mathcal{C} = 1.
\end{equation}
Our choices of $V$ and $Q$ for the other two symmetries with
\begin{equation}
V_\mathcal{T} = V_\mathcal{P} \equiv V, \quad Q_\mathcal{T} = Q_\mathcal{P} \equiv Q
\end{equation}
depend on the specific symmetry class.
When $\mathcal{P}^2 = -1$ (symmetry classes C, CI, CII) we choose $V = \sigma_y$, and we choose $V = 1$ in symmetry classes D, DIII, and BDI, where $\mathcal{P}^2 = 1$. The relative sign between $V$ and $Q$ follows from Eq.~\eqref{eq:vq_sign}:
\begin{equation}\label{eq:relation}
Q = \mathcal{P}^2 \mathcal{T}^2 V.
\end{equation}
In the remaining two classes AI and AII we choose $V=1$, and $Q=\mathcal{T}^2$. 
Symmetry representations of the effective Hamiltonians follow immediately from Eqs.~(\ref{eq:reducedham}).

Finally we show how symmetry operators change upon creating an effective Hamiltonian from a reflection matrix.
The effective Hamiltonian created from a reflection matrix with chiral symmetry satisfies
\begin{equation}
 H = \mathcal{P}^2 \mathcal{T}^2 V H^* V,
\end{equation}
so that the resulting symmetry of the effective Hamiltonian is particle-hole if $\mathcal{P}^2 \mathcal{T}^2 = -1$, and time-reversal if $\mathcal{P}^2 \mathcal{T}^2 = 1$. The symmetry operator of this symmetry squares to $VV^*$. If a reflection matrix has only time-reversal symmetry, then the time-reversal and particle-hole symmetry constraints on the effective Hamiltonian have the form
\begin{subequations}
 \begin{gather}
  H = \pm V \otimes \tau_x H^* V \otimes \tau_x,\\
  H = \mp V \otimes \tau_y H^* V \otimes \tau_y,
 \end{gather}
\end{subequations}
where the sign is determined by the choice of representation of the symmetry of $r$. Hence the symmetries of the effective Hamiltonian satisfy $\mathcal{T}^2 = -\mathcal{P}^2$. Finally, for an effective Hamiltonian constructed from a reflection matrix with only particle-hole symmetry, the resulting symmetry constraints on the Hamiltonian are
\begin{subequations}
 \begin{gather}
  H = \pm V \otimes \tau_0 H^* V \otimes \tau_0,\\
  H = \mp V \otimes \tau_z H^* V \otimes \tau_z,
 \end{gather}
\end{subequations}
so that both symmetry operators square to the same value.

\section{Calculation of the number of poles of \texorpdfstring{$\bm{\det r(z)}$}{det r(z)} inside the unit circle}\label{app:poles}

We prove that the equation
\begin{equation}
 \det[A-Z_1(k)] = 0
\end{equation}
has $N_1$ solutions with $z=e^{ik}$ inside the unit circle, and $N_1$
solutions outside of the unit circle as long as $A^\dagger A$ only has
eigenvalues less than one, which is generically the case since $A$ is
a sub-block of a unitary matrix $S$.  Let us assume that $\psi$ is an
eigenvector of the corresponding eigenvalue problem:
\begin{equation}
 A \psi = Z_1(k) \psi,
\end{equation}
with an eigenvalue with $|z|=1$. In this case $Z^\dagger_1(k) Z_1(k) =
1$. We come to a contradiction by considering the following inequality:
\begin{equation}
\psi^\dagger \psi > \psi^\dagger A^\dagger A \psi = \psi^\dagger
Z^\dagger_1(k) Z_1(k) \psi = \psi^\dagger \psi
\end{equation}
So we conclude that there are no solutions of $\det[A - Z_1(k)] = 0$ with
$z$ on the unit circle.  Next, we observe that for $A=0$ there are exactly
$N_1$ of $\det[A -Z_1(k)] = 0$ with $z = 0$ and $N_1$ solutions with $z =
\infty$. Since these solutions never cross the unit circle when $A$ is
smoothly deformed, we come to the statement we set to prove.

\bibliography{gtc}
\end{document}